\documentclass[12pt]{article}
\usepackage{amsmath,amssymb,amscd,cite,color}
\newtheorem{thm}{Theorem}[section]
\newtheorem{prop}[thm]{Proposition}
\newtheorem{lem}[thm]{Lemma}
\newtheorem{cor}[thm]{Corollary}

\newtheorem{defi}[thm]{Definition}
\newcommand{\pf}{{\bf Proof. \ }}
\newcommand{\qed}{\hfill $\blacksquare$ \\}

\font\msbm=msbm10 at 12pt

\newcommand{\F}{\mbox{\msbm F}}
\newtheorem{rem}[thm]{Remark}
\newtheorem{ex}[thm]{Example}

\begin{document}
\title{Constructions of Good Entanglement-Assisted Quantum Error Correcting Codes}

\author{Kenza Guenda, Somphong Jitman and T. Aaron Gulliver \thanks{K. Guenda
is with the Faculty of Mathematics USTHB, University of Science and
Technology of Algiers, Algeria.
S. Jitman  is with the Department of Mathematics, Faculty of Science, Silpakorn University,
Nakhon Pathom 73000, Thailand
T. A. Gulliver is with the Department of Electrical and Computer Engineering, University of
Victoria, PO Box 1700, STN CSC, Victoria, BC, Canada V8W 2Y2
email: kguenda@usthb.dz, jitmans@silpakorn.edu,agullive@ece.uvic.ca.}}

\maketitle

\begin{abstract}
Entanglement-assisted quantum error correcting codes (EAQECCs) are a simple and fundamental class of codes.
They allow for the construction of quantum codes from classical codes by relaxing the duality condition
and using pre-shared entanglement between the sender and receiver.
However, in general it is not easy to determine the number of shared pairs required to construct an EAQECC.
In this paper, we show that this number is related to the hull of the classical code.
Using this fact, we give methods to construct EAQECCs requiring desirable amount of entanglement. This leads to design families of EAQECCs with good error performance. Moreover, we construct maximal entanglement EAQECCs from LCD codes.
Finally, we prove the existence of asymptotically good EAQECCs in the odd characteristic case.
\end{abstract}

\section{Introduction}
Quantum codes are used to reduce decoherence over quantum information channels.
Several constructions for these codes have been proposed, the most important of which is the
CSS construction \cite{calderbank96,steane96} which provides stabilizer codes by exploiting the link between classical and quantum codes.
Other constructions of good quantum codes from classical codes include
the operator quantum error-correcting codes (OQECCs) introduced by Krib et al. \cite{krib}.
Although the OQECC construction provides good codes, the performance of the quantum system cannot be predicted from the properties of the underlying classical codes.
A simple and fundamental class of quantum codes called entanglement-assisted quantum error correcting codes (EAQECCs) was introduced by
Hsieh et al. \cite{hsieh}.
These codes have the advantages of both entanglement-assisted and operator quantum error correction.
They also showed that it is possible to construct entanglement-assisted operator quantum error correcting codes (EAOQECCs) from EAQECCs,
and in some cases EAQECCs can be used to obtain catalytic codes \cite{brun2}.
EAQECCs allow the use of arbitrary classical codes (not necessarily self-orthogonal) for quantum data transmission via pre-shared entanglement bits (ebits).
Further, the performance of the resulting quantum codes is determined by the performance of the underlying classical codes.
Fujiwara et al. \cite{fujwara} gave a general method for constructing entanglement-assisted quantum low-density parity check (LDPC) codes.
Hsieh et al. \cite{Hsieh} constructed EAQECC QC-LDPC codes which require only a small amount of initial shared entanglement.
Fan, Chen and Xu \cite{fan} provided a construction of entanglement-assisted quantum maximum distance separable (MDS)
codes with a small number of pre-shared maximally entangled states.
In addition, Qian and Zhang \cite{qian} constructed maximal-entanglement EAQECCs and proved the existence of asymptotically good EAQECCs in the binary case.

In this paper, good entanglement-assisted quantum codes are constructed.
First, a link between the number of maximally shared qubits required to construct an EAQECC from a classical code and the hull of the classical code is given.
Further, we give methods to construct EAQECCs requiring desirable amounts of entanglement.
This gives code designers flexibility in the choice of parameters, e.g. MDS or near MDS EAQECCs with a small number of pre-shared maximally entangled states.
These codes differ from those given in \cite{fan}.
In addition, EAQECCs are obtained from Reed-Solomon (RS) and generalized Reed-Solomon (GRS) codes.
Codes based on linear codes with complementary dual (LCD) are also given which
give rise to so-called maximal-entanglement EAQECCs introduced by Lai et al. \cite{wilde3}.
It was shown in \cite{wilde3} that maximal-entanglement EAQECCs are close to the hashing bound.
Motivated by this fact we construct EAQECC from LCD codes, further we prove the existence of a family of good EAQECCs from LCD codes.
LCD codes are also useful in that they provide flexibility in the choice of code parameters and can easily be decoded as shown by Massey \cite{massey}.

The remainder of this paper is organized as follows.
In Section 2 we provide some definitions and preliminary results.
In Section 3 we prove that the number of maximally entangled states is related to the hull of the classical codes.
Several constructions of EAQECCs with good performance and also with few shared states are presented in Section 4.
In Section 5 EAQECCs are constructed from linear codes with complementary dual (LCD).
Some of these codes are MDS.
Finally, an asymptotically good family of EAQECCs is obtained for the odd characteristic case.

\section{Preliminaries}
Let $\F_q$ denote the finite field of $q$ elements, where $q$ is a prime power.
For positive integers $k\leq n$ and $d$,
an $[n,k,d]_q $ {\em linear code} is defined to be a $k$-dimensional subspace of $\F_q^{n}$ with minimum Hamming distance $d$.
An $[n,k,d]_q$ code is called {\em maximum distance separable} (MDS) if the parameters satisfy $d=n-k+1$.

Let $\bar{} : \mathbb{F}_{q^2} \to \mathbb{F}_{q^2}$ be the map defined by $\overline{a}:=a^q$ for all $a\in \mathbb{F}_{q^2}$.
For a $k\times n$ matrix $A=(a_{ij})_{k\times n}$ and a vector $v=(v_1,v_2,\dots, v_n)$ over $\mathbb{F}_{q^2}$ (viewed as a $1\times n$ matrix),
let $\overline{A}:= (\overline{a_{ij}})_{k\times n}$ and $\overline{v}:=(\overline{v_1},\overline{v_2},\dots, \overline{v_n})$.
Denote by $A^\dagger$ and $v^\dagger$ the transpose matrices of $\overline{A}$ and $\overline{v}$, respectively.
For $v=(v_1 \ldots v_n)$ and $w=(w_1 \ldots w_n)$ in $\F_q^n$, the {\em  Euclidean
inner product} is defined by $\langle {v},{w}\rangle := \sum v_iw_i$, and
the {\em Hermitian inner product} is defined by $[{v},{w}] := \sum v_i\overline{w_i} $.
The {\em Euclidean} and {\em Hermitian dual codes} of $C$ are defined as
\[
C^\perp:=\{ {v} \in \F_q^n \ | \ \langle {v},{w}\rangle = 0 {\rm \ for\ all \ } {w} \in C\},
\]
and
\[ C^{\perp h}:=\{ {v} \in \F_{q^{2}}^n \ | \ [{v}, {w}]= 0 {\rm \ for\ all\ } {w} \in C\}.
\]

A linear code $C$ of length $n$ over $\F_q$ is said to be {\em cyclic} if it satisfies
\[
(c_{n-1},c_{0},\ldots,c_{n-2})\in C,\text{ whenever }(c_{0},c_{1},
\ldots, c_{n-1}) \in C.
\]
Further, a cyclic code of length $n$ is generated by a monic polynomial $g(x)$ which divides $x^n-1$.
Let $\alpha$ be a primitive $n$th root of unity in some extension field of $\mathbb{F}_q$.
The set $T$ of all integers $0\leq i<n$ such that $\alpha^i$ is a root of of $g(x)$ is called the {\em defining set} of $C$.
For $a\in\{0, \ldots, n-1\}$, the set $\{aq^j \bmod n \mid 0 \leq j < m \}$ is called a \emph{cyclotomic coset modulo} $n$ containing $a$.
It is well known that a defining set of a cyclic code of length $n$ is a union of cyclotomic cosets modulo $n$.
A polynomial $g(x)$ of degree $r$ over $\F_q$ with $g(0)\ne 0$ is called a {\em self-reciprocal polynomial} if $g(x)=g(0)^{-1}x^rg(x^{-1})$.

Generalized Reed-Solomon (GRS) codes are good codes for constructing EAQECCs.
The {\em GRS codes} are defined follows.
Let $\ell$ be a prime power.
For each positive integer $n\leq \ell $,
let $\gamma:=(\gamma_{1},\gamma_2,\ldots,\gamma_{n})$ and $w=(w_{1},w_2,\ldots,w_{n})$ where $\gamma_i$ is a non-zero element
and $w_{1},w_2,\ldots w_n$ are distinct elements in $\mathbb{F}_\ell$.
For each $0\leq k\leq n$, denote by $\mathbb{F}_\ell[X]_{k}$ the set of all polynomials of degree less
than $k$ over $\mathbb{F}_\ell$ (for convenience, the degree of the zero polynomial is defined to be $-1$).
A GRS code of length $n \leq q$ and dimension $k \leq n$ is defined as
 \begin{align}\label{def:GRS}
 {GRS}_{n,k}(\gamma,w):=
 \left\{(\gamma_{1}f(w_{1}),\gamma_2f(w_2),\ldots,\gamma_{n}f(v_{n})) \mid f(X) \in \mathbb{F}_\ell[X]_{k}\right\}\text{.}
 \end{align}
Choose the standard basis $\{1,x,\ldots,x^{k-1}\}$ for $\mathbb{F}_\ell[X]_{k}$.
A generator matrix of ${GRS}_{n,k}(w,\gamma)$ is given by
 \begin{equation}\label{eq:genGRS}
 G=\left(
 \begin{array}{cccc}
 \gamma_{1}            & \gamma_{2}            & \ldots & \gamma_{n}\\
 \gamma_{1}w_{1}  & \gamma_{2}w_{2}  & \ldots & \gamma_{n}w_{n}\\
 \vdots           & \vdots           & \ddots & \vdots \\
 \gamma_{1}w_{1}^{k-1}  & \gamma_{2}w_{2}^{k-1}  & \ldots & \gamma_{n}w_{n}^{k-1}
 \end{array}
 \right).
 \end{equation}
It is well known that ${GRS}_{n,k}(w,\gamma)$ is an MDS code with parameters $[n,k,n-k+1]_{\ell}$ and the Hermitian dual
$({GRS}_{n,k}(w,\gamma))^{\bot h}$ of ${GRS}_{n,k}(w,\gamma)$ is also a GRS code
${GRS}_{n,n-k}(v,\beta)$ for some $\beta,v\in \mathbb{F}_{q^2}^n$.

An $[[n,k,d;c]]_q$ entanglement-assisted quantum error-correcting code (EAQECC) encodes $k$ logical qudits into $n$ physical qudits using
$c$ copies of maximally entangled states.
The performance of an EAQECC is measured by its rate $\frac{k}{n}$ and net rate ($\frac{k-c}{n})$.
When the net rate of an EAQECC is positive it is possible to obtain catalytic codes as shown by Brun et al. \cite{brun2}.
In \cite{Wilde}, Wilde and Brun determined the optimal number of shared qubits.
In particular, they showed that EAQECCs can be constructed using classical linear codes as follows.

\begin{prop}[{\cite[Corollary 1]{Wilde}}]
\label{prop:ent1}
Let $H_1$ and $H_2$ be parity check matrices of two linear codes $[n,k_1,d_1]_{q}$ and $[n,k_2,d_2]_{q}$, respectively.
Then an $[[ n,k_1 +k_2-n+c, \min \{d_1,d_2\};c ]]_q$ EAQECC can be obtained where $c=rank (H_1H_2{^{t}})$ is the required number of maximally entangled states.
\end{prop}

It is also possible to construct EAQECCs in the Hermitian case using the following result.
\begin{prop}[{\cite[Corollary 2]{Wilde}}]
\label{prop:ent}
Let $H$ be the parity check matrix of an $[n,k,d]_{q^2}$ linear code over $\F_{q^2}$.
Then an $[[n,2k-n+c,d;c]]_{q}$ EAQECC can be obtained where $c=rank (HH^{\dagger})$ is the required number of maximally entangled states.
\end{prop}

An $[[n,2k-n+c,d;c]]_{q}$ EAQECC such that $c=n-k$ is called a {\em maximal-entanglement EAQECC}.
The Singleton bound for an EAQECC is given in the following proposition.
\begin{prop}[{\cite{brun}}]
\label{Bound:Singleton}
An $[[n,k,d;c]]_q$ EAQECC satisfies
\[
n +c -k \ge 2 (d-1),
\]
where $0 \le c \le n-1$.
\end{prop}
An EAQECC attaining this Singleton bound is called an {\em MDS EAQECC}.


 \section{The Number of Maximally Entangled States}

In this section, the problem of constructing EAQECCs with good performance is reduced
to finding classical codes with good error capability and also with large $rank(HH^{t})$ or $rank(HH^{\dagger})$.
For this, we provide a link between the number of maximally entangled states given by $rank(HH^t)$ (resp., $rank(HH^{\dagger})$) and the hull of a classical code.
 \subsection{The Euclidean Case}
We now provide a means of finding $rank(HH^t)$.
Let $C$ be a linear $[n,k,d]_q$ code with parity check matrix $H$.
Denote by $Hull(C)$ the {\em Euclidean hull} $C\cap C^{\bot} $ of $C$.
In the following proposition, we show that $rank(HH^t)$ is independent of $H$ and can be determined in terms of $Hull(C)$.

\begin{prop}
\label{prop:hull1}
Let $C$ be a linear $[n,k,d]_{q}$ code with parity check matrix $H$ and generator matrix $G$.
Then $rank(HH^t)$ and $rank(GG^t)$ are independent of $H$ and $G$ so that
\[
rank(HH^t)=n-k-\dim(Hull(C)) = n-k-\dim(Hull(C^{\bot })),
\]
and
\[
rank(GG^t)=k-\dim(Hull(C)) = k-\dim(Hull(C^{\bot })).
\]
\end{prop}
\pf Since $Hull(C)=Hull(C^{\bot })$, the second equality is obvious.
Let $m=\dim(Hull(C))$ and $B=\{h_1, h_2,\dots, h_m\}$ be a basis of $Hull(C)$.
Extend $B$ to be a basis
$\{h_1,h_2,$ $\dots,h_m,$ $h_{m+1},\dots, h_{n-k}\}$ of $C^{\bot }$.
Then
\[
K=\left(
\begin{array}{c} h_1\\h_2\\
\vdots \\
h_{n-k}
\end{array}
\right),
\]
is a parity check matrix of $C$.
Applying a suitable sequence of elementary row operations,
we have that $H=AK$ for some invertible $(n-k)\times (n-k) $ matrix $A$ over $\mathbb{F}_{q}$, and therefore
\[
HH^t= AK (AK)^t=AKK^t A^t.
\]
Since $A$ and $A^t$ are invertible, we have
\begin{align*}
    rank(HH^t)&=rank(K{K}^t)\\
    &=n-k-m\\
    &=n-k-\dim(Hull(C))\\
    &=n-k-\dim(Hull(C^{\bot })),
    \end{align*}
which is independent of $H$ as required.
Since $G$ is a parity check of $C^{\bot}$, a similar argument gives that $rank(GG^t)=k-\dim(Hull(C)) = k-\dim(Hull(C^{\bot }))$.
\qed

The following corollary is a direct consequence of Propositions \ref{prop:ent1}, \ref{Bound:Singleton} and \ref{prop:hull1}.
\begin{cor}
\label{cor:hull1}
Let $C$ be a classical $[n,k,d]_{q}$ linear code and $C^{\bot}$ its Euclidean dual with parameters $[n,n-k,d^{\bot} ]_{q}$.
Then there exist $[[n,k-\dim(Hull(C)), d;n-k-\dim(Hull(C))]]_{q}$ and $[[n,n-k-\dim(Hull(C)), d^{\bot};k-\dim(Hull(C))]]_{q}$ EAQECCs.
Further, if $C$ is MDS then the two EAQECCs are also MDS.
\end{cor}

\subsection{The Hermitian Case}
For a linear code $C$ over $\mathbb{F}_{q^2}$ with parity check matrix $H$,
denote by $Hull_h(C)$ the {\em Hermitian hull} $C\cap C^{\bot h} $ of $C$.
We show in the following proposition that $rank(HH^\dagger)$ is independent of $H$ and can be determined in terms of $Hull_h(C)$.

\begin{prop}
\label{prop:hull}
Let $C$ be a classical $[n,k,d]_{q^2}$ code with parity check matrix $H$ and generator matrix $G$.
Then $rank(HH^\dagger)$ and $rank(GG^\dagger))$ are independent of $H$ and $G$ so that
\[
rank(HH^\dagger)=n-k-\dim(Hull_h(C)) = n-k-\dim(Hull_h(C^{\bot h})),
\]
and
\[
rank(GG^\dagger)=k-\dim(Hull_h(C)) = k-\dim(Hull_h(C^{\bot h})).
\]
\end{prop}
\pf
Since $Hull_h(C)=Hull_h(C^{\bot h})$, the second equality is obvious.
Let $m=\dim(Hull_h(C))$ and $B=\{h_1, h_2,\dots, h_m\}$ be a basis of $Hull_h(C)$.
Extend $B$ to be a basis $\{h_1,h_2,$ $\dots,h_m,h_{m+1},$ $\dots, h_{n-k}\}$ of $C^{\bot h}$.
Let
\[
K=\left(\begin{array}{c} h_1\\h_2\\
    \vdots \\
    h_{n-k}
    \end{array}\right),
\]
so $\overline{K}$ is a parity check matrix of $C$.
After a suitable sequence of elementary row operations, we have that $H=A\overline{K}$ for some invertible $(n-k)\times (n-k) $ matrix $A$ over $\mathbb{F}_{q^2}$,
and then
\[
HH^\dagger= A\overline{K} (A\overline{K}) ^\dagger=A\overline{K}\,\overline{K}^\dagger A^\dagger.
\]
    Since $A$ and $A^\dagger$ are invertible, we have  \begin{align*}
    rank(HH^\dagger)&= rank(\overline{K}\,\overline{K}^\dagger)\\
    &= rank(K{K}^\dagger)\\
    &=n-k-m\\
    &=n-k-\dim(Hull_h(C))\\
    &=n-k-\dim(Hull_h(C^{\bot h}))
    \end{align*}
which is independent of $H$ as required.
Since $G$ is a parity check of of $C^{\bot}$, a similar argument gives that $rank(GG^\dagger)=k-\dim(Hull(C)) = k-\dim(Hull(C^{\bot h }))$.
\qed

The following corollary is a direct consequence of Propositions \ref{prop:ent}, \ref{Bound:Singleton} and \ref{prop:hull}.
\begin{cor}
\label{cor:hull}
Let $C$ be a classical $[n,k,d]_{q^2}$ code and let $C^{\bot h}$ be
its Hermitian dual with parameters $[n,n-k,d^{\bot h} ]_{q}$.
Then there exists $[[n,k-\dim(Hull_h(C)), d;n-k-\dim(Hull_h(C))]]_{q^2}$ and $[[n,n-k-\dim(Hull_h(C)), d^{\bot};k-\dim(Hull_h(C))]]_{q}$ EAQECCs.
If $C$ is MDS, then the two EAQECCs are also MDS.
\end{cor}




\section{The New Constructions}
In this section, we give some constructions of EAQECCs with few shared pairs.
Some of the resulting codes are MDS.
\subsection{ The Euclidean Case}
Two constructions of EAQECCs based on the Euclidean duals of linear codes are given below.

%
%
\begin{prop}
\label{propCons1}
Let $q>3$ be a prime power and let $C$ be a classical $[n,k,d]_{q} $ code such that $C^{\bot} \subseteq C$ and
$\dim(C) -\dim( C^{\bot })=\ell$.
Then for each $0\leq c\leq \ell$, there exists an $[[n+c,2k-n,d^\prime ;c]]_{q}$ EAQECC with $d\leq d^\prime \leq d+c$.
\end{prop}
\pf
Let $H$ be a parity check matrix for $C$ and let $D$ be a linear code such that $C^\perp \oplus D=C$.
Further, let $x_1, x_2,\dots, x_c$ be linearly independent codewords in $D$.
Moreover, $x_1, x_2,\dots, x_c$ can be chosen such that $ x_i x_i^t \neq 0$ and $ x_i x_j^t =0 $ for all $1\leq i<j\leq c$.
Since $q>3$ and $\{a^2\mid a\in \mathbb{F}_q^*\}$ contains at least $2$ elements,
for each $i\in \{1,2,\dots,c\}$ there exists $\alpha_i\in \mathbb{F}_{q}^*$
such that $\alpha_i^2\ne -x_ix_i^t$.
Note that the $\alpha_i$ are not necessarily distinct.
Let $C^\prime$ be the code with parity check matrix
\[
H'=\left(\begin{array}{ccc|c}0&&&H\\ \hline
 \alpha_1&&  &x_1\\
     &\ddots&&\vdots\\
    & &\alpha_c&x_c
    \end{array}\right).
\]
Since $\alpha_i\ne -x_ix_i^t$ for all $1\leq i \leq c$, we have that $rank(H^\prime (H^\prime)^t)=c$.
Further, as every $d-1$ columns of $H$ are linearly independent and $\alpha_i\ne 0$ for all $i\in \{1,2,\dots,c\}$,
every $d-1$ columns of $H^\prime$ are linearly independent.
It follows that $C$ is an $[n+1,k, d^\prime]_{q}$ code where $d\leq d^\prime\leq d+c$.
Then by Proposition \ref{prop:ent1}, there exists an $[[n+c,2k-n,d^\prime ;c]]_{q}$ EAQECC.
\qed

\begin{ex}
An excellent family of classical codes to obtain EAQECCs using the proposed construction is the class of Reed-Solomon (RS)codes.
Recall that an RS code denoted $\mathcal{RS}_{n,k}$ is a cyclic MDS codes of length $n:=q-1$ over $\F_q$ with generator polynomial
$g(x)= (x-\alpha) \ldots (x-\alpha^{r-1})$ and parameters $[n,n-r+1,r]_q$, where $\alpha$ is a primitive element of $\mathbb{F}_q$.
In this case, each cyclotomic coset contains only one element.
The code $\mathcal{RS}_{n,k}^{\bot}$ is equal to $\mathcal{RS}_{n,n-k}$.
Hence if $n< 2k$ or equivalently $r < \frac{n+1}{2}$, then $\mathcal{RS}_{n,k}$ will be dual containing.
Thus is $T=\{1, \ldots, r\}$ is the defining set of $\mathcal{RS}_{n,k}$, then the dual code has defining set $T=\{1, \ldots, r-l\}$,
so from Proposition \ref{propCons1} there exists a $[[q+c-1,2k+1-q, d'\ge n-k+1;c]]_q$ code for all $ c \le l$.
\end{ex}
\begin{prop}
\label{prop:conJ}
Let $q$ be a prime power, $C$ be an $[n,k,d]_{q}$ code such that $C^{\bot} \subseteq C$, and
$c\leq n-k+1$ be a positive integer.
Then there exists an $[[n+1,2k-n-1+c,d^\prime;c]]_{q}$ EAQECC where $d^\prime\in\{d,d+1\}$ if one of the following conditions holds.
    \begin{enumerate}
        \item[(i)]  $q=2$ and $c$  is odd.
        \item[(ii)]   $q=3$  and $3\nmid c$.
       \item[(iii)]   $q\geq 4$.
       \end{enumerate}
\end{prop}
\pf Two cases need to be considered, 1) $\gcd(q,c)=1$, and 2) $q\geq 4$ and $\gcd(q,c)\ne 1$.
Let $x$ be an element in $\mathbb{F}_q^{n-k}$ defined by
\[x:=\begin{cases}
(0,0,\dots,0) &\text{ if } c=1,\\
(\underbrace{1,\dots,1}_{c-1 \text{ copies}}, 0,\dots,0) &\text{ if } 2\leq c\leq n-k+1.
\end{cases}\]
Then there exists $a\in \mathbb{F}_q\setminus \{-1\}$ such that
\begin{align*}
 xx^t=c-1&= \begin{cases}
a\ne -1& \text{ if }   \gcd(q,c)= 1,\\
-1 & \text{ if }    \gcd(q,c)\ne 1.
\end{cases}
\end{align*}
Let $\omega$ be a primitive element of $\mathbb{F}_q$ and let $\alpha$ be an element of $\mathbb{F}_q$ defined by
 \[ \alpha := \begin{cases}
 1 & \text{ if }   \gcd(q,c)= 1,\\
 \omega  & \text{ if }   q\geq 4 \text{ and }  \gcd(q,c)\ne 1.
 \end{cases}\]
 Since $\omega^2\ne 1$ for all $q\geq 4$, it follows that $xx^t\ne -\alpha^2$

Without loss of generality, assume that $H=(I_{n-k} \,\, A)$ is a parity check matrix of $C$.
 Let $C^\prime $ be the linear code with parity check matrix
\[H'=\left(\begin{array}{ccc}
\alpha &x&0\\
0&I_{n-k}&A
\end{array}\right).\]
Since
\[
H^\prime(H^\prime)^t=\left(\begin{array}{ccc}
\alpha^2+xx^t&x&0\\
x^t&I_{c-1}&0 \\
0&0&0
\end{array}\right),
\]
and $-xx^t\ne \alpha^2$, we have $rank(H^\prime(H^\prime)^t)=c$.
It is not difficult to determine that every $d-1$ columns of $H^\prime$ are linearly independent.
Hence $C^\prime$ is an $[n+1,k,d^\prime]_q$ code with $d^\prime\in\{d,d+1\}$.
Then by Proposition \ref{prop:ent1}, there exists an $[[n+1,2k-n-1+c,d^\prime ;c]]_{q}$ EAQECC.
\qed

\begin{rem} From the well-known CSS construction \cite{calderbank96,steane96}
of symmetric quantum codes based on Euclidean dual-containing codes,
an $[[n,2k-n,d]]_q$ CSS code can be constructed if and only if there exists a Euclidean dual-containing $[n,k,d]_q$ code.
Then combined with Propositions \ref{propCons1} and \ref{prop:conJ},
it can be concluded that if there exists an $[[n,2k-n,d]]_q$ CSS code, then
EAQECCs with the following parameters can be constructed
\begin{enumerate}
\item[i)] $[[n+c,2k-n,d';c]]_q$ with $d'\geq d$ for all $0\leq c\leq n-k$, and
\item[ii)] $[[n+1,2k-n-1+c,d';c]]_q$ with $d'\geq d$ for all $1\leq c\leq n-k+1$.
\end{enumerate}
Therefore, many EAQECCs can be constructed from Propositions \ref{propCons1} and \ref{prop:conJ}.
\end{rem}

\subsection{The Hermitian Case}

In this subsection, we construct EAQECCs based on Hermitian dual-containing classical linear codes.
We first extend the Euclidean constructions given previously to the Hermitian case.

%
%
\begin{prop}
\label{prop:herm1}
Let $q>2$ be a prime power and $C$ be an $[n,k,d]_{q^2}$ code such that $C^{\bot h} \subseteq C$ and $\dim(C) -\dim( C^{\bot h})=\ell$.
Then for each $0\leq c\leq \ell$, there exists an $[[n+c,2k-n,d^\prime ;c]]_{q}$ EAQECC with $d\leq d^\prime \leq d+c$.
\end{prop}
\pf Let $H$ be a generator matrix for $C^{\bot h}$, $D$ be a linear code such that $C^{\perp h} \oplus D=C$, and
$x_1, x_2,\dots, x_c$ be linearly independent codewords in $D$.
Moreover, $x_1, x_2,\dots, x_c$ can be chosen such that $ x_i x_i^\dagger \neq 0$ and $ x_i x_j^\dagger =0$
for all $1\leq i<j\leq c$.
For each $i\in \{1,2,\dots,c\}$, there exist $\alpha_i\in \mathbb{F}_{q^2}^*$ such that $\alpha_i{^{q+1}}\ne -x_ix_i^\dagger$.
Let $C^\prime$ be the code with parity check matrix
\[
H'=\left(\begin{array}{ccc|c}0&&&H\\ \hline
 \alpha_1&&  &x_1\\
      &\ddots&&\vdots\\
    & &\alpha_c&x_c
    \end{array}\right).
 \]
Since $\alpha_i^{q+1}\ne -x_ix_i^\dagger$ for all $1\leq i \leq c$, we have that $rank(H^\prime (H^\prime)^\dagger)=c$.
As every $d-1$ columns of $H$ are linearly independent and $\alpha_i\ne 0$ for all $i\in \{1,2,\dots,c\}$,
every $d-1$ columns of $H^\prime$ are linearly independent.
It follows that $C$ is an $[n+1,k, d^\prime]_{q^2}$ code where $d\leq d^\prime\leq d+c$, and then
by Proposition \ref{prop:ent} there exists an $[[n+c,2k-n,d^\prime ;c]]_{q}$ EAQECC.
\qed

In the following proposition, MDS EAQECCs are obtained using the construction given in Proposition \ref{prop:herm1}
and the dual-containing GRS codes defined in \eqref{def:GRS}.

\begin{prop}
\label{prop:classicMDS}
Let $q> 2$ be a prime power and $1\leq n \leq q^2$ be an integer.
Further, let $C$ be an $[n,k,n-k+1]_{q^2}$ Hermitian dual-containing GRS code.
$C^{\bot h} $ is generated by
\[
H=\left(
	\begin{array}{cccc}
	\beta_{1}            & \beta_{2}            & \ldots & \beta_{n}\\
	\beta_{1}v_{1}  & \beta_{2}  v_{2}& \ldots & \beta_{n}v_{n}\\
	\vdots           & \vdots           & \ddots & \vdots \\
	\beta_{1}v_{1}^{n-k-1}  & \beta_{2}v_{2}^{n-k-1}  & \ldots & \beta_{n}v_{n}^{n-k-1}
	\end{array}
	\right),
\]
for some non-zero $\beta_i$ and distinct elements $v_i$ in $\mathbb{F}_{q^2}$.
If $x=(\beta_{1}v_{1}^{n-k}, \beta_{2}v_{2}^{n-k}, \ldots, \beta_{n}v_{n}^{n-k})$
and $\alpha\in \mathbb{F}_{q^2}^*$ such that $\alpha^{q+1}\ne -xx^\dagger$, then
\[
H'=\left(\begin{array}{cc}0&H\\
	\alpha &x\\
	\end{array}\right),
\]
is a parity check matrix of an $[n+1,k,n-k+2]_{q^2}$ MDS code with $rank(H^\prime (H^\prime)^\dagger)=1$.
In this case,
$[[n+1,2k-n,n-k+2;1]]_{q}$ and $[[n+1,1,k+1,2k-n-1]]_{q}$ MDS EAQECCs can be constructed.
\end{prop}
\pf Let $C^\prime$ be a linear code with parity check matrix $H^\prime$.
Then by Proposition \ref{prop:herm1},
$C^\prime$ is an $[n+1,k,d]_{q^2}$ code with $n-k+1\leq d\leq n-k+2$ and $rank(H^\prime (H^\prime)^\dagger)=1$.
Since the code with parity check matrix
$\left(\begin{array}{c}H\\ x\\ \end{array}\right)$
is GRS, $C^\prime $ is an extended GRS code which is MDS.
Hence, $C^\prime$ is an $[n+1,k,n-k+2]_{q^2}$ MDS code, and
an $[[n+1,2k-n,n-k+2;1]]_{q}$ MDS EAQECC exists by Proposition \ref{prop:herm1}.

By Proposition \ref{prop:hull}, $(C^\prime)^{\bot h}$ is an $[n+1,n-k+1,k+1] _{q^2}$ code with $\dim(Hull_h(C^\prime))=n-k$.
Hence there exists an $[[n+1,1,k+1;2k-n-1]]_q$ MDS EAQECC by Corollary \ref{cor:hull}.
\qed

From Proposition \ref{prop:classicMDS}, an MDS EAQECC can be constructed whenever a Hermitian dual-containing (or equivalently self-orthogonal)
GRS code exists.
Hermitian dual-containing GRS codes have been extensively studied, e.g. \cite{JLLC2010,ZG2015}.
For the parameters given in Table \ref{table:mds1}, there exists an $[n,k,n-k+1]_{q^2}$ Hermitian dual-containing GRS code (see the corresponding references).
Then by Proposition \ref{prop:classicMDS}, there exists an $[n+1,k,n-k+2]_{q^2}$ code $C$ with $\dim(Hull_h(C))=n-k$, so
$[[n+1,2k-n,n-k+2;1]]_{q}$ and $[[n+1,1,k+1;2k-n-1]]_{q}$ MDS EAQECCs can be constructed.

\begin{table}[!hbt] \centering
	\begin{tabular}{|c|c|c|l|}
		\hline
		$q$&$n$& $k$  & Reference\\
		\hline
		arbitrary  &$rm\leq n\leq rm+1$,  &$1\leq k\leq \frac{m-1}{q+1}$&   \cite[Theorem 2.3]{JLLC2010} \\
		& $m|(q^2-1)$  and && \\
		&  $0\leq r\leq \frac{q^2-1}{m}$&& \\ \hline
		arbitrary     &$mq-q+1\leq n\leq mq$,&$ n-\frac{(q-1-\lfloor r/m\rfloor)}{2}\leq k \leq n-2$&  \cite[Theorem  3.4]{JLLC2010} \\
		&$1\leq m\leq q$&& \\
		\hline
		$q=2am+ 1$ &$\frac{q^2-1}{a}$&$n- (a + 1)m\leq k \leq n-1$&  \cite[Theorem 3.2]{ZG2015} \\ \hline
		$q=2am- 1$ &$\frac{q^2-1}{2a}-q+1$&$  n- (a + 1)m + 3\leq k\leq n-1$&\cite[Theorem 3.7]{ZG2015} \\
		\hline
	\end{tabular}
	\caption{Parameters for Constructing MDS EAQECCs}
	\label{table:mds1}
\end{table}

\begin{prop}
\label{prop:conJ2}
Let $q>2$ be a prime power, $C$ be an $[n,k,d]_{q^2}$ code such that $C^{\bot h} \subseteq C$, and $c\leq n-k+1$ be a positive integer.
Then there exists an $[[n+1,2k-n-1+c,d^\prime;c]]_{q}$ EAQECC where $d^\prime\in\{d,d+1\}$.
\end{prop}
\pf
Let $x$ be an element in $\mathbb{F}_{q^2}^{n-k}$ defined by
\[
x:=\begin{cases}
(0,0,\dots,0) &\text{ if } c=1,\\
(\underbrace{1,\dots,1}_{c-1 \text{ copies}}, 0,\dots,0) &\text{ if } 2\leq c\leq n-k+1.
\end{cases}
\]
Then there exists $a\in \mathbb{F}_{q^2}\setminus \{-1\}$ such that
\begin{align*}
xx^\dagger=c-1&= \begin{cases}
        a\ne -1& \text{ if }   \gcd(q,c)= 1,\\
        -1 & \text{ if }    \gcd(q,c)\ne 1.
        \end{cases}
\end{align*}
Let $\omega$ be a primitive element of $\mathbb{F}_{q^2}$ and $\alpha$ be an element of $\mathbb{F}_{q^2}$ defined by
\[
        \alpha := \begin{cases}
        1 & \text{ if }   \gcd(q,c)= 1,\\
        \omega  & \text{ if }    \gcd(q,c)\ne 1.
        \end{cases}
\]
Since $\omega^{q+1}\ne 1$, it follows that $xx^\dagger\ne -\alpha^{q+1}$.
Without loss of generality, assume that $H=(I_{n-k} \,\, A)$ is a generator matrix for $C^{\bot h}$.
Let $C^\prime $ be the code with parity check matrix
\[
        H'=\left(\begin{array}{ccc}
        \alpha &x&0\\
        0&I_{n-k}&A
        \end{array}\right).
\]
Since
\[
H^\prime(H^\prime)^\dagger=\left(\begin{array}{ccc}
        \alpha^{q+1}+xx^\dagger&x&0\\
        x^\dagger&I_{c-1}&0 \\
        0&0&0
        \end{array}\right),
\]
and $-xx^\dagger \ne \alpha^{q+1}$, $rank(H^\prime(H^\prime)^\dagger)=c$.
It is not difficult to determine that every $d-1$ columns of $H^\prime$ are linearly independent.
Hence $C^\prime$ is an $[n+1,k,d^\prime]_{q^2}$ code with $d^\prime\in\{d,d+1\} $.
Then by Proposition \ref{prop:ent}, there exists an $[[n+1,2k-n-1+c,d^\prime ;c]]_{q}$ EAQECC.
\qed

\begin{rem}
From the well-known CSS construction \cite{calderbank96,steane96,JLLC2010} of symmetric quantum codes based on Hermitian dual-containing codes,
an $[[n,2k-n,d]]_q$ CSS code can be constructed if and only if there exists a Hermitian dual-containing $[n,k,d]_{q^2}$ code.
Then with Propositions \ref{prop:herm1} and \ref{prop:conJ2},
it can be concluded that if there exists an $[[n,k,d]]_q$ CSS code,
then EAQECCs with the following parameters can be constructed:
\begin{enumerate}
\item[i)] $[[n+c,2k-n,d';c]]_q$  with $d'\geq d$ for all $0\leq c\leq n-k$, and
\item[ii)] $[[n+1,2k-n-1+c,d';c]]_q$  with $d'\geq d$ for all $1\leq c\leq n-k+1$.
\end{enumerate}
Therefore many EAQECCs can be constructed from Propositions \ref{prop:herm1} and \ref{prop:conJ2}.
\end{rem}

%

\subsection{MDS EAQECCs from the Hermitian Hulls of GRS Codes}

In this section, a construction of MDS EAQECCs is presented which is based on the dimension of the Hermitian hull of GRS codes.
In order to determined $Hull_h({GRS}_{n,k}(\gamma,w))$,
we begin with the following lemma regarding finite fields.

\begin{lem}
\label{sum:0}
Let $\ell$ be a prime power and $i\geq 0$ be an integer.
Then $\sum\limits_{a\in\mathbb{F}_{\ell}^*} a^{i} =0$ if and only if $(\ell-1)\nmid i$.
\end{lem}
\pf If $(\ell-1) | i$, then $a^i=1$ for all $a\in\mathbb{F}_{\ell}^*$, and then
$\sum\limits_{a\in\mathbb{F}_{\ell}^*} a^{i} =\ell-1\ne 0\in \mathbb{F}_{\ell}$.
Conversely, assume that $(\ell-1)\nmid i$.
If $\omega$ is a primitive element of $\mathbb{F}_{\ell}$, then
$\omega^i\ne 1$ and $(\omega^i)^{\ell-1}=1$.
Hence $\sum\limits_{a\in\mathbb{F}_{\ell}^*} a^{i} =\sum\limits_{ j=0}^{l-2}(\omega^i)^j =((\omega^i)^{\ell-1}-1)(\omega^i-1)^{-1}=0$ as required.
\qed

The dimension of the Hermitian hull of some GRS codes is determined in the following proposition.

\begin{prop}
\label{prop:hullGRS}
Let $q>2$ be a prime power, $n\in \{ (q-1)r, (q-1)r+1\mid 1\leq r\leq q+1$, and $\gcd(r,q)=1\}$.
Then there exist distinct elements $\gamma\in (\mathbb{F}_{q^2}^*)^n$ and $w\in \mathbb{F}_{q^2}^n$ such that:
\begin{enumerate}
            \item[(i)]  $\dim(Hull_h(GRS_{(n,0)}(\gamma,w))) =0$, and
            \item[(ii)] for  each $1\leq k\leq n$,  $(i-1)(q-1)< k\leq i (q-1)$ for some positive integer $i$  and
             \begin{align*}\dim(&Hull_h(GRS_{(n,k)}(\gamma,w)))\\
            &=\begin{cases}
            \dim(Hull_h(GRS_{(n,k-1)}(\gamma,w))) &\text{ if }  k=(i-1)(q-1),\\
            \dim(Hull_h(GRS_{(n,k-1)}(\gamma,w))) -1
            &\text{ if }   (i-1)(q-1)+1< k\leq q(i-1)+i+1,\\
            \dim(Hull_h(GRS_{(n,k-1)}(\gamma,w)))+1 &\text{ if }  q(i-1)+i+1<k\leq i(q-1).
            \end{cases}
            \end{align*}
\end{enumerate}
\end{prop}
\pf
Let $\omega$ be a primitive element of $\mathbb{F}_q$ and
$\{\beta_0=1, \beta_1,\beta_2,\dots,\beta_q\}$ be a complete set of representatives of the cosets of the multiplicative group $\mathbb{F}_q^*$ in $\mathbb{F}_{q^2}^*$.
First consider the case $n\in \{ (q-1)r\mid 1\leq r\leq q+1$ and $\gcd(r,q)=1\}$.
For each $1\leq r \leq q$, let
\[
w:=(\beta_0,\beta_0\omega,\dots,\beta_0\omega^{q-2}, \beta_1,\beta_1\omega,\dots,\beta_1\omega^{q-2}, \dots , \beta_{r-1},\beta_{r-1}\omega,\dots,\beta_{r-1}\omega^{q-2})
\]
and let $\gamma:=(1,1,\dots,1)\in \mathbb{F}_{q^2}^n$.
Then the elements in $w$ are distinct.

The first statement is obvious.
To prove the second statement, assume that $1\leq k\leq n$.
Clearly, $(i-1)(q-1)< k\leq i (q-1)$ for some positive integer $i$.
For convenience, denote by $g_j$ the $j$th row of the generator matrix of $GRS_{(n,k)}(\gamma,w)$ as given in \eqref{eq:genGRS}.
Consider the following three cases.
\begin{itemize}
\item[Case 1:] $k=(i-1)(q-1)+1$.
Then $(q^2-1)|(k-1+q(k-1))$ and $(q^2-1)\nmid (k-1+qj)$ for all $0\leq j< k-1$.
It follows from Lemma \ref{sum:0} that
\begin{align}
\label{eq:innerprod}
g_kg_{j+1}^\dagger=\sum_{t=0}^{r-1}\left(\beta_t^{k-1+qj}\sum_{m=0}^{q-2} \omega ^{m(k-1+qj)}\right)
=\sum_{t=0} ^{r-1} \left(\beta_t^{k-1+qj}\cdot 0 \right)=0,
\end{align}
for all $0\leq j< k-1$ and $g_kg_k^\dagger\neq 0\in \mathbb{F}_{q^2}$.
Consequently
\[
Hull_h(GRS_{(n,k)}(\gamma,w))= Hull_h(GRS_{(n,k-1)}(\gamma,w)).
\]

\item[Case 2:] $(i-1)(q-1)+1< k\leq q(i-1)+i+1$.
Then there exists a unique positive integer $s<k-1$ such that $(q^2-1)|(k-1+sq)$.
Similar to \eqref{eq:innerprod}, it follows from Lemma \ref{sum:0} that $g_kg_{j+1}^\dagger=0$ for all $0\leq j<s$ and $s<j\leq k-1$,
and $g_k g_{s+1}^\dagger\neq 0$.
We have that $Hull_h(GRS_{(n,k-1)}(\gamma,w)) = Hull_h(GRS_{(n,k)}(\gamma,w))\oplus \langle g_{s+1}\rangle$, and hence
\[
\dim(Hull_h(GRS_{(n,k)}(\gamma,w))) = \dim(Hull_h(GRS_{(n,k-1)}(\gamma,w))) -1.
\]

\item[Case 3:] $q(i-1)+i+1<k\leq i(q-1)$.
In this case, there are no integers $s\leq k-1$ such that $(q^2-1)| (k-1+qi)$.
Similar to \eqref{eq:innerprod}, we have that $g_kg_j^\dagger= 0$ for all $1\leq j \leq k$.
It follows that $Hull_h(GRS_{(n,k)}(\gamma,w)) = Hull_h(GRS_{(n,k-1)}(\gamma,w))\oplus \langle g_k\rangle$, and hence
\[
\dim(Hull_h(GRS_{(n,k)}(\gamma,w))) = \dim(Hull_h(GRS_{(n,k-1)}(\gamma,w))) +1.
\]
\end{itemize}
We now consider $n\in \{ (q-1)r+1\mid 1\leq r\leq q+1 \text{ and } \gcd(r,q)=1\}$.
In this case, there exists $\alpha\in \mathbb{F}_{q^2}^*$ such that $\alpha^{q+1}\ne-\gamma\gamma^\dagger$.
Let $\gamma^\prime=(\alpha,1,1,\dots, 1)$ and
\[
w^\prime=(0,1,\omega,\omega^2,\dots,\omega^{q-2}, \beta_1,\beta_1\omega,\beta_1\omega^2,\dots,\beta_1\omega^{q-2}, \dots ,\beta_{r-1},\beta_{r-1}\omega,\beta_{r-1}\omega^2,\dots,\beta_{r-1}\omega^{q-2}).
\]
Using arguments similar to the previous case, it can be shown that $GRS_{(n,k)}(\gamma^\prime,w^\prime)$ has the required properties.
\qed

From Proposition \ref{prop:hullGRS}, the dimension of the Hermitian hull of $GRS_{(n,k)}(\gamma,w)$ can be determined recursively on $k$.
Therefore, MDS EAQECCs corresponding to these codes can be constructed.

Using the fact that $Hull_h(GRS_{(n,k)}(\gamma,w))=Hull_h((GRS_{(n,k)}(\gamma,w))^{\bot h})$,
Proposition \ref{prop:hullGRS} and Corollary \ref{cor:hull},
some parameters can be explicitly stated as in the following corollaries.

\begin{cor}
Let $q>2$ be a prime power, $n\in \{ (q-1)r, (q-1)r+1\mid 1\leq r\leq q+1 \text{ and } \gcd(n,q)=1\}$, and $ 1\leq k <q-1$.
Then there exist $[n,k,n-k+1]_{q^2}$ and $[n,n-k,k+1]_{q^2}$ MDS codes such that $\dim(Hull_h(C))=k-1$, so
there exist $[[n,1,n-k+1;n-2k+1]]_q$ and $[[n, n-2k +1 ,k+1;1]]_q$ MDS EAQECCs.
\end{cor}

\begin{cor}
Let $q>2$ be a prime power, $n\in \{ (q-1)r, (q-1)r+1\mid 1\leq r\leq q+1 \text{ and } \gcd(n,q)=1\}$, and $ q-1\leq k< 2(q-1)$.
Then there exist $[n,k,n-k+1]_{q^2}$ and $[n,n-k,k+1]_{q^2}$ MDS codes such that $\dim(Hull_h(C))=k-2$, so
there exist $[[n,2,n-k+1;n-2k+2]]_q$ and $[[n, n-2k +2 ,k+1;2]]_q$ MDS EAQECCs.
\end{cor}

\begin{cor}
Let $q>$ be a prime power, $n\in \{ (q-1)r, (q-1)r+1\mid 1\leq r\leq q+1 \text{ and }\gcd(n,q)=1\}$, and $k=2(q-1)$.
Then there exist $[n,k,n-k+1]_{q^2}$ and $[n,n-k,k+1]_{q^2}$ MDS codes such that $\dim(Hull_h(C))=k-3$, so
there exist $[[n,3,n-k+1;n-2k+3]]_q$ and $[[n, n-2k +3,k+1;3]]_q$ MDS EAQECCs.
\end{cor}

\section{EAQECCs from LCD codes}

{\em Linear codes with complementary dual} (LCD) are defined to be linear codes $C$ whose dual codes
$ C^{\bot } $ satisfy $ C \cap C^{\bot }=\lbrace0\rbrace $\cite{massey}.
In this section, we construct EAQECCs from LCD codes.
We have the following result from \cite{massey} which is a corollary of Proposition \ref{prop:hull}.

\begin{prop}
 \label{lem:2}
If $H$ is a parity check matrix of an $[n,k]_q$ linear code $C$, then $C$ is an
LCD code if and only if the $(n-k) \times (n-k)$ matrix $HH^t$ is nonsingular.
\end{prop}

It is obvious that if $C$ is an $[n,k,d]_q$ LCD code, then its dual is an $[n,n-k, d^{\bot}]_q$ LCD code.
From Proposition \ref{prop:hull}, it can be determined that the largest entanglement occurs with LCD codes.
Using Corollary \ref{cor:hull1}, we obtain the following result.

\begin{prop}
\label{prop:cons}
If there exists an $[n,k,d]_q$ LCD code $C$, then there exist $[[n,k,d,n-k]]_q$ and $[[n,n-k,d^\perp,k]]_q$ maximal-entanglement EAQECCs
where $d^\perp$ is the minimum distance of $C^\perp$.
 \end{prop}

In \cite{massey2}, the following result was given concerning cyclic LCD codes.
\begin{lem}
\label{lem:massey2}
Assuming that $(n,q)=1$, if $g(x)$ is the generator polynomial of an $[n,k,d]_q$ cyclic code $C$,
then $C$ is an LCD code if and only if $g(x)$ is a self-reciprocal polynomial.
\end{lem}

We now give  an infinite family of maximal-entanglement EAQECCs which are also MDS.

\begin{thm}
If $q$ is even, then there exists MDS maximal-entangled EAQECC with parameters $[[q+ 1, k, q-k+ 2, q+1-k]]_q$ for all integers $k$ such that $1 \le k \le q + 1$.
If $q$ is odd, then there exists MDS maximal-entangled EAQECC with parameters $[[q+ 1, k, q-k+ 2, q+1-k]]_q$ for all odd integers $k$ such that $1 \le k \le q + 1$.
\end{thm}
\pf
From \cite[Theorem 8]{grassl}, if $q+1-k$ is odd (this case correspond to $q$ and $k$ both even or odd),
then the cyclic code generated by the polynomial $g_1(x)= \prod_{i=- \mu}^\mu(z-\alpha ^{i})$ is a
$[ q+1,q-2\mu, 2\mu+2]_q$ MDS cyclic code.
Since $g_1(x)$ is self-reciprocal, the codes are LCD.
The results then follow from Proposition \ref{prop:cons}.

If $q$ is even and $k$ is odd, then the polynomial $g_2(x)= \prod_{i=q/2- \mu}^{q/2}(z- \alpha ^{i})(z-\alpha ^{-i}) $
generates a $[ q+1,q-1-2\mu, 2\mu+3]_q$ MDS cyclic code from \cite[Theorem 8]{grassl}.
Since $g_2(x)$ is self-reciprocal, the codes are LCD by Lemma \ref{lem:massey2}.
The results then follow from Proposition \ref{prop:cons}.
\qed

\begin{thm}
\label{thm:LCD}
Assume that $q= p^r$ is a prime power integer.
If an $[n,k,d]_q$ linear code over $\F_q$ exists, then there exists
an $[[N, k,d';c]]_q$ EAQECC with $ sd -1 \ge d' \ge d$ and $(N,c)$ as follows:
\begin{enumerate}
\item[(i)] $(N,c)=(2n-k, 2n-2k)$ if $q$ is even and $s=2$,
\item[(ii)] $(N,c)=(3n-2k,3n-3k) $ if $q \equiv 1 \bmod 4$ and $s=3$,
\item[(iii)] $(N,c)=(4n-3k, 4n-4k))$ if $q \equiv 3 \bmod 4$ and $s=4$, and
\item[(iv)] $(N,c)=(5n-4k, 5n-5k)$ for any $q $ and $s=5$.
\end{enumerate}
\end{thm}
\pf
Let $C$ be a linear code with parameters $[n,k,d]_q$ and generator matrix $G= (I_k \ A)$.
For even $q$ and $s=2$, let $C'$ be a linear code with generator matrix $G'=(I_k \ A \ A)$.
A simple calculation shows $G'(G')^t=I_k$.
Hence, $C'$ is  a $[2n-k,k, d' \ge d]_q$ code with parity check matrix $H'$ such that $rank(H'(H')^t)= 2n-2k$, and
therefore, there exists a $[[2n-k,k,d' \ge d;2n-2k]]_q$ EAQECC.

If $q \equiv 1 \bmod 4$ and $s=3$, then there exists $\alpha \in \F_q$ such that $\alpha^2+1=0$.
The matrix $G'= (I_k\ A \ \alpha A)$ generates an LCD code $C'$ over $\F_q$ with parity check matrix $H'$ such that $rank(H(H')^t)= 3n-3k$.
Hence from Proposition \ref{prop:cons} there exists a $[[3n-2k,k, d' \ge d; 3n-3k]]_q$ EAQECC.

If $q\equiv 3 \bmod 4$ and $s=4$, then from \cite[p. 281]{rosen} there exist $\alpha, \beta \in \F_q$ such that $\alpha^2 +\beta^2+1=0$.
Hence the matrix $G'= (I _n\ A\ \alpha A\ \beta A)$ generates a $[4n-3k,k, d' \ge d]_q$ LCD code $C'$ over $\F_q$ with parity check matrix $H'$ such that $rank(H(H')^t)= 4n-4k$.
Therefore from Proposition \ref{prop:cons} there exists a $[[4n-3k,k, d'\ge d ;4n-4k]]_q$ EAQECC.

If $q$ is a prime power and $s=5$, then from \cite[Theorem 370]{hardy} we have that every prime is the sum of four squares.
Then there exist $\alpha, \beta, \gamma$ and $\delta$ in $\mathbb{F}_q$
such that $\alpha^2+ \beta^2+\gamma^2+\delta^2 =p$, and
the matrix $G'= (I_n\ A \alpha A\ \beta A \ \delta A\ \gamma A)$ generates an LCD code over $\F_q$ with parity check matrix $H'$ such that $rank(H(H')^t)= 5n-5k$.
Hence from Proposition \ref{prop:cons} there exists a $[[5n-4k, k,5n-5k ,d';5n-5k]]_q$ EAQECC with $ d' \ge d$.

Finally, if $G$ contains a codeword of minimum weight, then in each construction above $d'\leq sd-1$.
\qed

One may ask if the EAQECCs obtained in Proposition \ref{thm:LCD} are good, i.e., if they have good rate and positive net rate.
A simple calculation gives the following results.
\begin{cor}
\label{cor:LCD1}
If an $[n,k,d]_q$ linear code exists, then from Theorem \ref{thm:LCD} there exists an $[[N, k,d';c]]_q$ LCD EAQECC
with positive net rate and rate larger than $1/2$ if we have the following:
\begin{enumerate}
\item[(i)] $k/n > 2/3$ if $q$ is even,
\item[(ii)] $k/n > 3/4$ if $q \equiv 1 \bmod 4$, or
\item[(iii)] $k/n > 4/5$ if $q \equiv 3 \bmod 4$.
\end{enumerate}
\end{cor}
%
%
%
%

\subsection{Asymptotically Good EAQECCs}
Qian and Zhang \cite{qian} used binary LCD codes which are transitive to prove the existence of an asymptotically good family of EAQECCs \cite{stichtenoth}.
We prove in this section that the same arguments are valid for finite fields of odd characteristic.

\begin{defi}
Let $\mathcal{C}$ be a family of $ [n_i, k_i, d_i]_q$ linear codes.
Then $\mathcal{C}$ is called {\em asymptotically good} if $R > 0$ and $\delta > 0$ where $R$ is the {\em  asymptotic rate} of $\mathcal{C}$ defined as
$R = \lim_{ i \rightarrow \infty } \frac{k_i}{n_i}$ and $\delta $ is the {\em relative distance} of $\mathcal{C}$
defined as $\delta := \lim_{i \rightarrow \infty } \frac{d_i}{n_i}$.
\end{defi}

\begin{defi}
\label{codeexpa}
Let $C$ be an $[n, k, d_1]_{q^{m}}$ code over $\F_{q^{m}}$ and
$\beta := \{b_1 , \ldots , b_m \}$ be a basis of $\F_{q^{m}}$ over $\F_q$.
Then the $q$-ary expansion of $C$ with respect to $\beta$,
denoted by $\beta (C)$, is a linear $q$-ary code with parameters
$[nm, mk, d_2 \geq d_1]_{q}$ given by
$\beta (C):= \{{(c_{ij})}_{i,j} \in \F_{q}^{mn}| (c_1,c_2, \ldots, c_n) \in C \text{ and } c_i = \displaystyle\sum_{j}c_{ij}b_{j}\}$.
\end{defi}

A subgroup $\mathcal{G}$ of the symmetric group $S_n$ is called {\em transitive} if for any pair
$(i , j)$, $1\le i ,j \le n$, there exists a permutation $\sigma \in \mathcal{G}$ such that $\sigma(i) = j$.
A permutation $\sigma\in S_n $ is called an {\em automorphism} of the code $C \subseteq \F_q^{ n}$ provided that for each
vector $(c_1, \ldots, c_n) \in C$, the vector $(c_{\sigma(1)}, \ldots, c_{\sigma(n)})$ is also in $C$.
Then $Aut(C)$ is the group of all automorphisms of $C$.

\begin{defi} A code C over $\F_q$ of length $n$ is said to be transitive if its automorphism
group $Aut(C)$ is a transitive subgroup of $S_n$.
\end{defi}

Using the geometric Goppa codes, Stichtenoth \cite{stichtenoth} proved the following result.

\begin{thm}
\label{thm:stichtenoth}
Let $q = l^2$ and $R, \delta > 0$ be real numbers with $R = 1-\delta-1/(l-1)$.
Then there exists a sequence $(C_j)_{j\ge 0}$  of linear codes $C_j = [n _j , k _j , d_j ]_q $ with the following properties:
\begin{enumerate}
\item[(i)] $C_j$ is a transitive code,
\item[(ii)] $n_j \rightarrow \infty$ as $j \rightarrow \infty $, and
\item[(iii)] $ \lim_{ j \rightarrow \infty } \frac{k_j}{n_j} \ge R$ and $\lim_{ j \rightarrow \infty } \frac{d_j}{n_j} \ge \delta$.
\end{enumerate}
\end{thm}

Then we have the following result which gives an asymptotically good family of EAQECCs.

\begin{thm}
If $q=l^{2m}$, where $l$ is an odd prime, then there exists a family of EAQECCs $Q_j$ with parameters $[[n_j, k_j, d_j; c_j]]_q$
such that $\lim_{ j \rightarrow \infty } \frac{k_j}{n_j} > 0$ and $\lim_{ j \rightarrow \infty } \frac{d_j}{n_j} > 0$.
\end{thm}
\pf
Let $\mathcal{C} :=(C_j)_{j \ge 0}$ be the transitive family of cides in Theorem \ref{thm:stichtenoth}.
Then the code expansion $\beta (C_j)$ has parameters $[mn_j, mn_j, \ge d_j]_{l^2}$ over $\F_{l^2}$.
Since $l$ is odd, we have that $l^2 \equiv 1 \bmod 4$, and then by Theorem \ref{thm:LCD} there exists an $[[n_h, k_h, d_h; c_h]]_{l^2}$ EAQECC,
where $n_h= 3 m n_j - 2m k_j$, $k_h= m k_j$, $d_h \ge d_j,$ and $c_h= 3m n_j - 3m k_j$.
From Theorem \ref{thm:stichtenoth} it can be concluded that
\[
R=\lim \frac{k_j}{n_j} =\lim \frac{mk_j}{3mn_j-2mk_j} \ge \lim \frac{mk_j}{3n_j} >0,
\]
and
\[
\delta =\lim \frac{d_h}{n_h} \ge lim \frac{mk_j}{3mn_j-2mk_j} \ge \lim \frac{d_j}{3mn_j} > 0,
\]
as required.
\qed

 \end{document}